\documentclass[aps,prb,reprint]{revtex4-2}
\usepackage{graphicx} 
\usepackage{amsmath} 
\usepackage{bm} 
\usepackage{natbib}
\usepackage{natmove}
\usepackage[colorlinks=true, linkcolor=blue, citecolor=blue, urlcolor=blue, linktoc=page, bookmarks=false, pdfstartview={FitH}, pdfborder={0 0 0.0 [3 3]}]{hyperref} 
\usepackage{color} 
\usepackage{dcolumn} 
\newcolumntype{.}{D{.}{.}{0.3}}
\newcolumntype{-}{D{.}{.}{4.0}}
\usepackage{makecell}
\usepackage{multirow}
\usepackage{cleveref}
\usepackage{easyReview} 
\crefname{figure}{Fig.}{Figs}
\crefname{table}{Table}{Tables}
\crefname{section}{Sec.}{Secs.}
\crefname{equation}{Eq.}{Eqs.}
\usepackage[mathscr]{euscript}
\newcommand\identity{1\kern-0.25em\text{l}}
\usepackage{braket}
\renewcommand{\today}{\number\day \space \ifcase \month \or January\or February\or March\or April\or May\or June\or July\or August\or September\or October\or November\or December\fi \space \number\year} 
\def\m1r{\multicolumn{1}{r}}

\begin{document}
\title{Nonsymmorphic symmetry-enforced hourglass fermions and Rashba-Dresselhaus interaction in BiInO$_3$}
\author{Ramsamoj \surname{Kewat}}
\email[Email: ]{ramsamoj18@iiserb.ac.in}
\author{Nirmal \surname{Ganguli}}
\email[Email: ]{NGanguli@iiserb.ac.in}
\affiliation{Department of Physics, Indian Institute of Science Education and Research Bhopal, Bhauri, Bhopal 462066, India}
\date{\today}
\begin{abstract}
In this study, we investigate the spin texture of the hourglass fermions band network in BiInO$_3$ using density functional theory (DFT) and symmetry analysis. Hourglass fermions are of interest in spintronics due to their unique and robust band structure, as well as their potential applications in novel electronic devices. BiInO$_3$ exhibits non-symmorphic crystal symmetries, such as glide reflection and glide rotational symmetry, influencing its electronic properties. Through symmetry analysis, we explore the band crossings and spin textures along specific high-symmetry paths in the Brillouin zone. Our results reveal a fascinating hourglass-shaped band dispersion and spin polarisation governed by symmetry operations and spin-orbit interaction. We analyse the spin-splitting mechanisms, including Dresselhaus and Rashba spin-orbit interactions, and suggest potential applications for spin-based devices. This study sheds light on the role of symmetry in crystals for fascinating spin properties of hourglass fermions in non-symmorphic materials, offering insights for future developments in spintronics.
\end{abstract}

\maketitle

\section{Introduction}

In recent years, there has been a paradigm shift in electronic device fabrication, moving away from traditional charge-based systems towards the promising realm of Spintronics. Unlike conventional electronics that depend on manipulating electronic charges, Spintronics exploits the intrinsic quantum-mechanical property of electron spin. 
In Spintronics, materials with remarkable quantum effects, such as Spin-Orbit Interaction (SOI), symmetry, topology, and dimensionality, play essential roles. Topological insulators\cite{Kane2005, Moore2007, Fu2006, Chiu2016, Fu2007}, Hourglass fermions\cite{Wang2016b,Li2018, Wang2017,Wang2019}, Weyl semimetals\cite{PhysRevA.94.053619, Soluyanov2015,Zhang2017, Rosenstein2018,Yao2019}, and superconductors\cite{Chiu2016,PhysRevB.93.195413} are notable among these materials, offering unique opportunities for novel device architectures and functionalities. Spatial symmetry plays a crucial role in classifying materials with band topology. The crystal symmetry will differentiate the materials, whether it is symmorphic or non-symmorphic. Materials with only point group symmetry are symmorphic and those with glide reflection or screw rotation symmetry are non-symmorphic. The non-primitive translation along with rotational or mirror symmetry in non-symmorphic crystals leads to an hourglass cone-shaped band first predicted theoretically for the (010) surface of the KHgX(X=As, Sb, Bi) family \cite{Wang2016b} and then studied experimentally by angle-resolved photoemission Spectroscopy (ARPES)\cite{Ma2017}. However, later on, many researchers found several hourglass fermions in 3D and 2D materials\cite{Wang2017, Wang2019}. Non-symmorphic material inherently has bulk asymmetry, which opens the path to exploring interesting spin textures that might be present in them. The spin-orbit interaction (SOI) in condensed matter physics proposed by Dresselhaus\cite{1001954} and Rashba \cite{RashbaSPSS60, Rashba1960, Bychkov1984, Rashba1965} gives an insight into the novel properties of materials such as momentum-dependent band splitting and spin-momentum locking. Dresselhaus SOI manifests in materials lacking bulk inversion symmetry, giving rise to a unique momentum-dependent magnetic field $\vec{B}=\alpha(k_y,k_x)$ acting on electron spins. Conversely, Rashba SOI emerges at material surfaces or interfaces having a momentum-dependent magnetic field $\vec{B}=\beta(k_y,-k_x)$ on the electron. These interactions lead to a chiral spin texture in the electronic band structure, a crucial characteristic exploited in the design of spin field-effect transistors (spin FETs)\cite{Datta1990}.
The momentum-dependent shape of spin has some challenges and limitations in spintronics device applications. One of the challenging issues is related to spin lifetime, which is dictated by the scattering of spins due to impurities and defects within the material. This leads to rapid spin relaxation, known as Dyakonov-Perel spin relaxation\cite{Dyakonov2017}. This phenomenon significantly affects the practical realisation and application of spin FETs.
To overcome these limitations and unlock Spintronics's full potential, researchers actively seek materials exhibiting a non-chiral or momentum-independent spin texture in their electronic band structure. Such materials, characterised by equal Rashba and Dresselhaus SOI strengths, are termed Persistent Spin Textures (PSTs)\cite{Koralek2009, Schliemann2017,Bandyopadhyay2021}. By having unique properties of PST materials, it becomes possible to overcome the challenges associated with spin relaxation, paving the way for developing highly efficient and reliable spin-based electronic devices. While exploring non-symmorphic material, BiInO$_3$ is a good candidate for hourglass fermions with unique PST spin texture due to the presence of a glide mirror and glide rotational symmetry. The following section will discuss a detailed study of its band structure and spin texture using the DFT method and symmetry analysis.

\section{\label{sec:method}Crystal structure and Method}
\begin{figure*}[] \label{fig:1}
    \centering
    \includegraphics[scale=0.27]{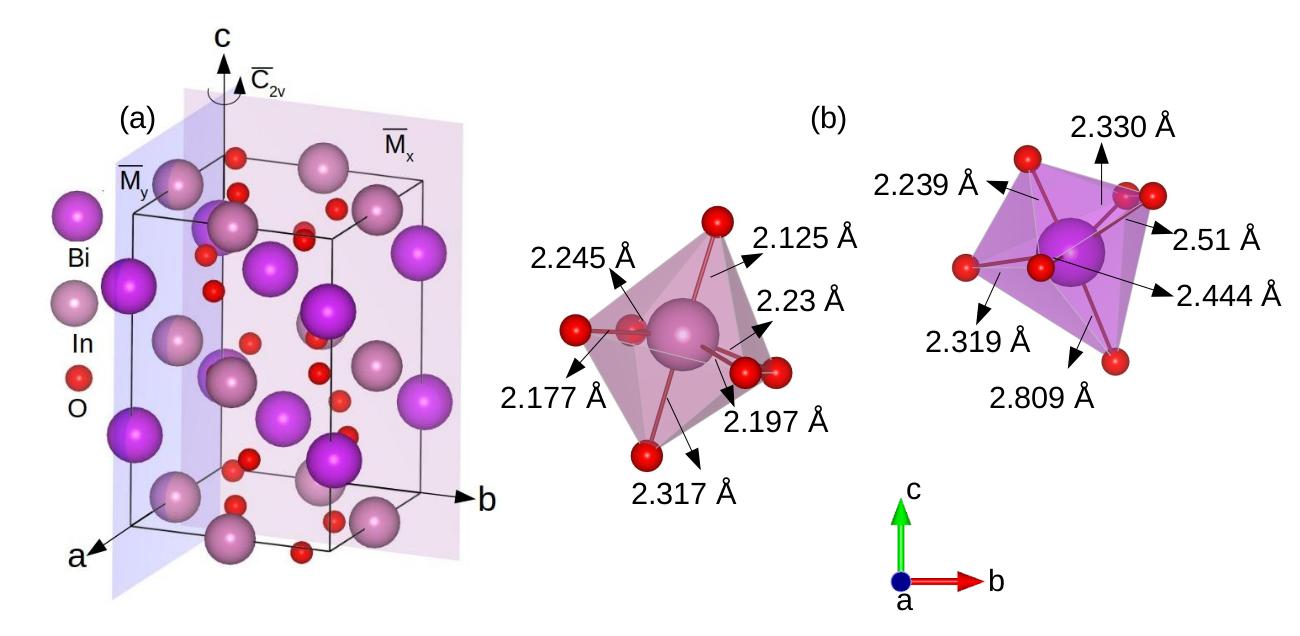}
    \caption{\label{fig:biin_plane}(a) Crystal structure of BiInO$_3$ with symmetry planes. $\Tilde{\mathcal{M}}_x$ is glide reflection plane about x= 0 axis. $\Tilde{\mathcal{M}}_y$ is glide reflecton plane about y=0 axis and $\Tilde{\mathcal{C}}_{2v}$ is screw rotational axis pointing along the c-axis. (b) The distorted In and Bi octahedra show the inequivalent bond length.}
\end{figure*}

BiInO$_3$ crystallises into a non-centrosymmetric polar orthorhombic crystal with space group $Pna2_1$ (\#33) \cite{Belik2006}. Besides the identity, the space group comprises only three symmetry operations, namely, a twofold screw rotation about the $z$-axis $\Tilde{\mathcal{C}}_{2z}$ (\{$2_{001}|0~0~\frac{1}{2}$\} in Seitz notation), a glide reflection about the $x = 0$ plane $\Tilde{\mathcal{M}}_x$ (\{$m_{100}|\frac{1}{2}~\frac{1}{2}~\frac{1}{2}$\} in Seitz notation), and a glide reflection about the $y = 0$ plane $\Tilde{\mathcal{M}}_y$ (\{$m_{010}|\frac{1}{2}~0~\frac{1}{2}$\} in Seitz notation), a glide reflection about the $x = 0$ plane $\Tilde{\mathcal{M}}_x$ (\{$m_{010}|\frac{1}{2}~\frac{1}{2}~0$\} in Seitz notation). The non-centrosymmetric polar space group allows for a substantial electric polarisation along the $z$-direction, assisted by the displacement of the heavy Bi and In ions and distorted Bi and In octahedra. The distortion of these octahedra results in an unequal bond length of Bi-O and In-O as shown in \cref{fig:biin_plane}. The Bi-O bond length ranges from 2.319\AA ~to 2.809\AA, and the In-O bond length ranges from 2.177\AA ~to 2.317\AA. The distorted octahedral and non-uniform bond length of Bi-O and In-O results in BiO and InO$_2$ plane displacement along the z-direction. The planar distance between BiO and InO$_2$ plane is alternately 2.345~\AA ~and 1.847~\AA. Since the BiO plane carries a unit positive charge and InO$_2$ carry a unit negative charge, the planar displacement causes an asymmetrical distribution of charges and results in an internal electric field along the z-direction.

The electronic structure calculations presented here are performed using density functional theory (DFT) within the framework of the projector augmented wave (PAW) method \cite{paw} along with a plane wave basis set, as implemented in the {\scshape vasp} code \cite{vasp1, vasp2}. A 500~eV plane wave energy cutoff is used for expanding the wavefunctions. An approximate form of the exchange-correlation functional through the generalised gradient approximation due to Perdew, Burke, and Ernzerhof (PBE) \cite{pbe} is used for the calculations. A $11\times13\times9$ $\Gamma-$centered k-mesh centred at the $\Gamma$-point is used for the Brillouin zone integration within the corrected tetrahedron method \cite{BlochlPRB94T} for the calculations, respectively. The crystal structure is optimised to find atomic positions with Hellmann-Feynman force on each atom below a threshold of $10^{-2}$~eV/\AA\ and the lattice vectors with minimum stress. The electric polarisation is calculated using the full potential linear augmented plane wave (FP-LAPW) method, as implemented in the WIEN2k code \cite{10.1063/1.5143061}, and the BerryPI code \cite{BerryPI} interfaced with the WIEN2k code.

\section{\label{sec:results}Results and discussions}

\subsection{Symmetries in BiInO$_3$}
\begin{figure*} [t!]
    \centering
  \includegraphics[scale=0.65]{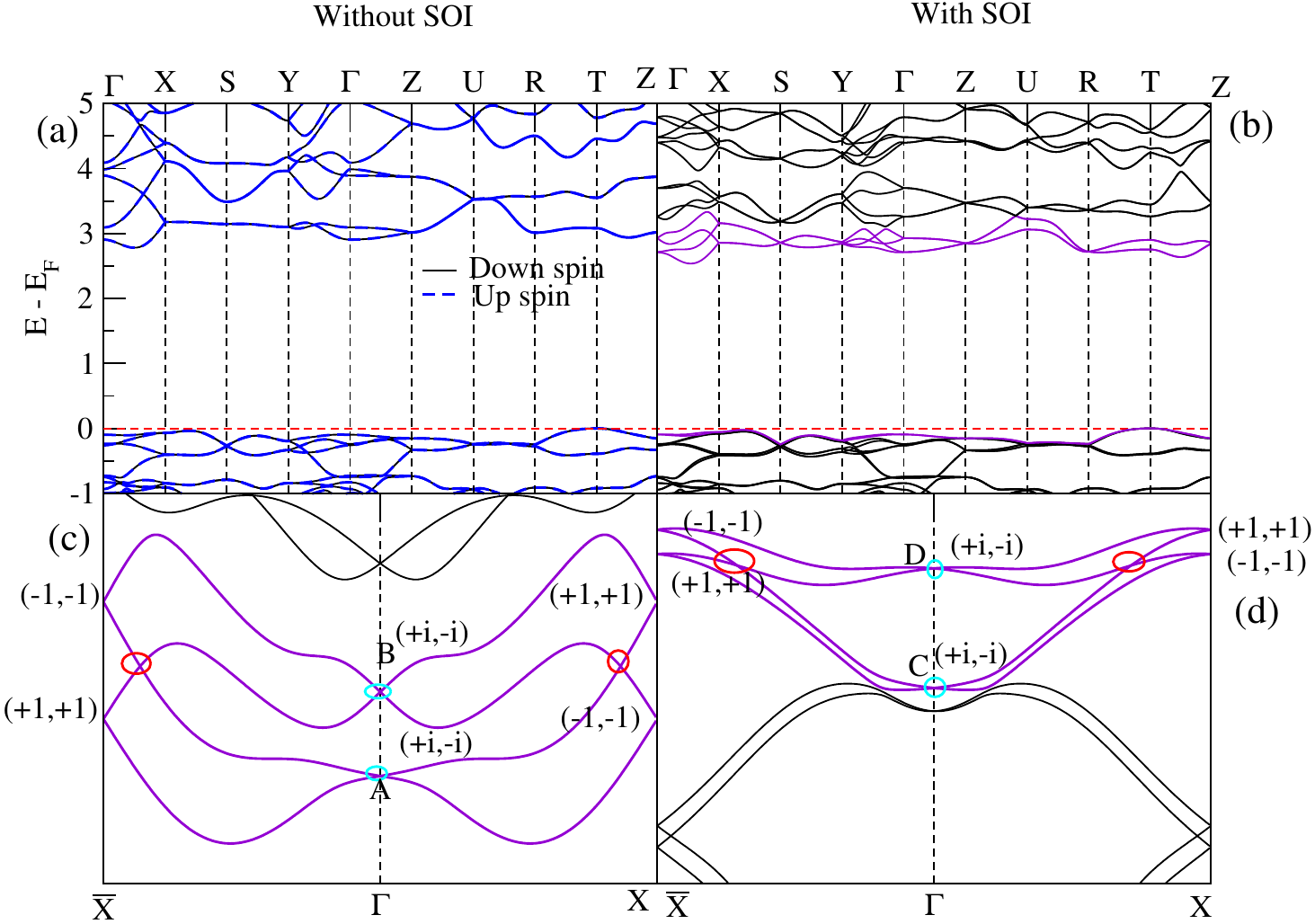}
   \caption{\label{fig:comb}Energy band structure and spin textures in BiInO$_3$.  
(a) and (b) show the bulk band structures of BiInO$_3$ without and with spin–orbit interaction (SOI), respectively.  
(c) illustrates the hourglass fermion network in the conduction band near the $\Gamma$ point. The red circles mark the hourglass crossings, while the cyan circles indicate time-reversal invariant momenta (TRIMs). Points A and B on the energy axis at $\Gamma$ represent states with a persistent spin texture pointing along the $k_y$ direction. These states begin with eigenvalues (+i, –i) and evolve toward the X point, ending with eigenvalues (+1, +1) and (–1, –1).  
(d) shows a similar hourglass fermion network in the valence band, close to the Fermi level, also centered at the $\Gamma$ point. Here, points C and D indicate states with a Rashba-like spin texture. These also begin with eigenvalues (+i, –i) and evolve toward X with final eigenvalues (+1, +1) and (–1, –1).}

\end{figure*}

Materials that exhibit symmetry-enforced band crossings possess distinctive characteristics within topological materials. Notably, these band crossings are persistent in the presence of specific symmetries inherent in the material, but they remain susceptible to movement under certain conditions. The mobility of these band crossings primarily arises from Nonsymmorphic crystal symmetries and non-spatial symmetries.

Nonsymmorphic crystal symmetry, denoted as $\mathcal{G}=\{g,t\}$, is characterized by a combination of point-group symmetries, such as reflection symmetry along the ($x=0$ and $y=0$) plane, rotational symmetry along the z-axis in the case of BiInO$_3$, and translation symmetry $T_a$ by a fraction of the Bravais lattice vector $\vec{a}$. Repeated application of an n-fold Nonsymmorphic symmetry results in the translation of Bloch states n-1 times, expressed as:

\begin{equation} \label{eq:1}
\mathcal{G}^n = \{g^n|nt\}=\pm \mathcal{P}T_a, \quad \mathcal{P}\in{1,2,......,n-1} 
\end{equation}

Here, $g$ represents the n-fold point-group symmetry, $T_a$ is the translation operator for the Bravais lattice vector, and the $\pm$ sign on the right-hand side signifies Boson and Fermion states, respectively.

BiInO$_3$ exemplifies a material with Pna2$_1$ orthorhombic perovskite structure, falling under space group No.33. This material manifests various symmetries, including the Identity symmetry (\identity) and Time reversal symmetry ($\mathcal{T}$), contributing to its unique topological characteristics. The following are the symmetries present in BiInO$_3$.

(1) Glide reflection Symmetry about yz-plane,

\begin{align} \label{eq:2}
\left\{ m_{100} \,\middle|\, \tfrac{1}{2}~\tfrac{1}{2}~\tfrac{1}{2} \right\}: (x, y, z) \rightarrow \left( -x + \tfrac{1}{2},\, y + \tfrac{1}{2},\, z + \tfrac{1}{2} \right)
\end{align}

(2) Glide reflection Symmetry about the xz-plane,
\begin{align} \label{eq:3}
\left\{ m_{010} \,\middle|\, \tfrac{1}{2}~0~\tfrac{1}{2} \right\}: (x, y, z) \rightarrow \left( x + \tfrac{1}{2},\, -y + \tfrac{1}{2},\, z \right)
\end{align}

(3) Two-fold screw rotation symmetry about the z-axis,
\begin{align} \label{eq:4}
\left\{ 2_{001} \,\middle|\, 0~0~\tfrac{1}{2} \right\}: (x, y, z) \rightarrow \left( -x,\, -y,\, z + \tfrac{1}{2} \right)
\end{align}

The crystallographic symmetry inherent in materials significantly influences their electronic structure. In the case of BiInO$_3$, the interplay between non-symmorphic symmetries and Spin-Orbit Interaction (SOI) is pivotal in determining the band structure, giving rise to phenomena such as band degeneracy and splitting. Our focus is understanding how specific symmetries, such as glide reflection symmetry and time reversal, contribute to forming Kramers-like degeneracy along distinct high symmetry paths in momentum space, as illustrated in \cref{fig:comb}(b). Before delving into the Eigenvalues and Eigenstates along these paths, it's necessary to comprehend how the glide reflection and time reversal operators act on the Bloch Eigenstates.
 
BiInO$_3$ is non-magnetic, preserving time inversion symmetry, which becomes a key aspect of our study. The time inversion operator, being antiunitary for Fermions, is a crucial element in understanding the band degeneracy and splitting along the chosen high symmetry path in the momentum space of this material. The time inversion operator is written as,

\begin{equation} \label{eq:5}
\mathcal{T}= \textit{i}\sigma_y\hat{\mathcal{K}}
\end{equation}
\begin{equation} \label{eq:6}
\mathcal{T}^2=\textit{i}^2\sigma_y^2\hat{\mathcal{K}}^2=-\identity
\end{equation}
where $\sigma_y$ Pauli spin matrix and $\hat{\mathcal{K}}$ is a complex conjugation operator in Bloch State.

The time inversion operator commutes with all the space group transformations and considers spin-orbit interaction and all the space group symmetry also present in spin space. Let us define the time-spatial symmetries $\mathcal{T}\Tilde{\mathcal{M}_x}=\mathcal{T}\left\{ m_{100} \mid \frac{1}{2}~\frac{1}{2}~\frac{1}{2} \right\}$ and $\mathcal{T}\Tilde{\mathcal{M}_y}=\mathcal{T}\left\{ m_{010} \mid \frac{1}{2}~0~\frac{1}{2} \right\}$ also preserved in the crystal and particular section of momentum space. The Glide reflection symmetry, Time inversion symmetry and their product can operate on Bloch states having both crystal space and spin space. Operating the symmetry operator on position  and spin space twice, we get the following equations,
\begin{equation} \label{eq:7}
   \{m_{100}^2|0~1~1\}:(x,y,z)\rightarrow (x,y+1,z+1)\textit{i}^2\sigma_x^2  
\end{equation}
\begin{equation} \label{eq:8}
    \{m_{100}^2|0~1~1\}:(x,y,z)\rightarrow - e^{-\textit{i}(k_y+k_z)}(x,y,z)
\end{equation}
\begin{equation} \label{eq:9}
   \{m_{010}^2|1~0~0\}:(x,y,z)\rightarrow (x+1,y,z)\textit{i}^2\sigma_y^2  
\end{equation} 
\begin{equation} \label{eq:10}
    \{m_{010}^2|1~0~0\}:(x,y,z)\rightarrow - e^{-\textit{i}k_x}(x,y,z)
\end{equation}
\begin{equation} \label{eq:11}
  \mathcal{T}^2\{m_{100}^2|0~1~1\}:(x,y,z)\rightarrow e^{-\textit{i}(k_y+k_z)}(x,y,z)
\end{equation}
\begin{equation} \label{eq:12}
  \mathcal{T}^2\{m_{010}^2|1~0~0\}:(x,y,z)\rightarrow e^{-\textit{i}k_x}(x,y,z)
\end{equation}
Let us redefine $\psi$ to the Bloch state. From \cref{eq:7,eq:8,eq:9,eq:10,eq:11,eq:12} we can get eigenvalue equation of the symmetry operators as;
\begin{equation} \label{eq:13}
\left\{ m_{100} \mid \frac{1}{2}~\frac{1}{2}~\frac{1}{2} \right\}\psi^{\pm}(\vec{k})=\pm\textit{i}e^{-\textit{i}(k_y+k_z)/2}\psi^{\pm}(\vec{k})
\end{equation}
\begin{equation}\label{eq:14}
\left\{ m_{010} \mid \frac{1}{2}~0~\frac{1}{2} \right\}\psi^{\pm}(\vec{k})=\pm\textit{i}e^{-\textit{i}k_x/2}\psi^{\pm}(\vec{k})
\end{equation}
\begin{equation}\label{eq:15}
\mathcal{T}\left\{ m_{100} \mid \frac{1}{2}~\frac{1}{2}~\frac{1}{2} \right\}\psi^{\pm}(\vec{k})=\pm e^{-\textit{i}(k_y+k_z)/2}\psi^{\pm}(\vec{k})
\end{equation}
\begin{equation}\label{eq:16}
\mathcal{T}\left\{ m_{010} \mid \frac{1}{2}~0~\frac{1}{2} \right\}\psi^{\pm}(\vec{k})=\pm e^{-\textit{i}k_x/2}\psi^{\pm}(\vec{k})
\end{equation}
where eigenvalues are momentum-dependent and can be calculated to some constant, by knowing the value of momentum at a high symmetry point or path.
\begin{figure*}
\centering
  \graphicspath{ {/}}
   \includegraphics[scale=0.51]{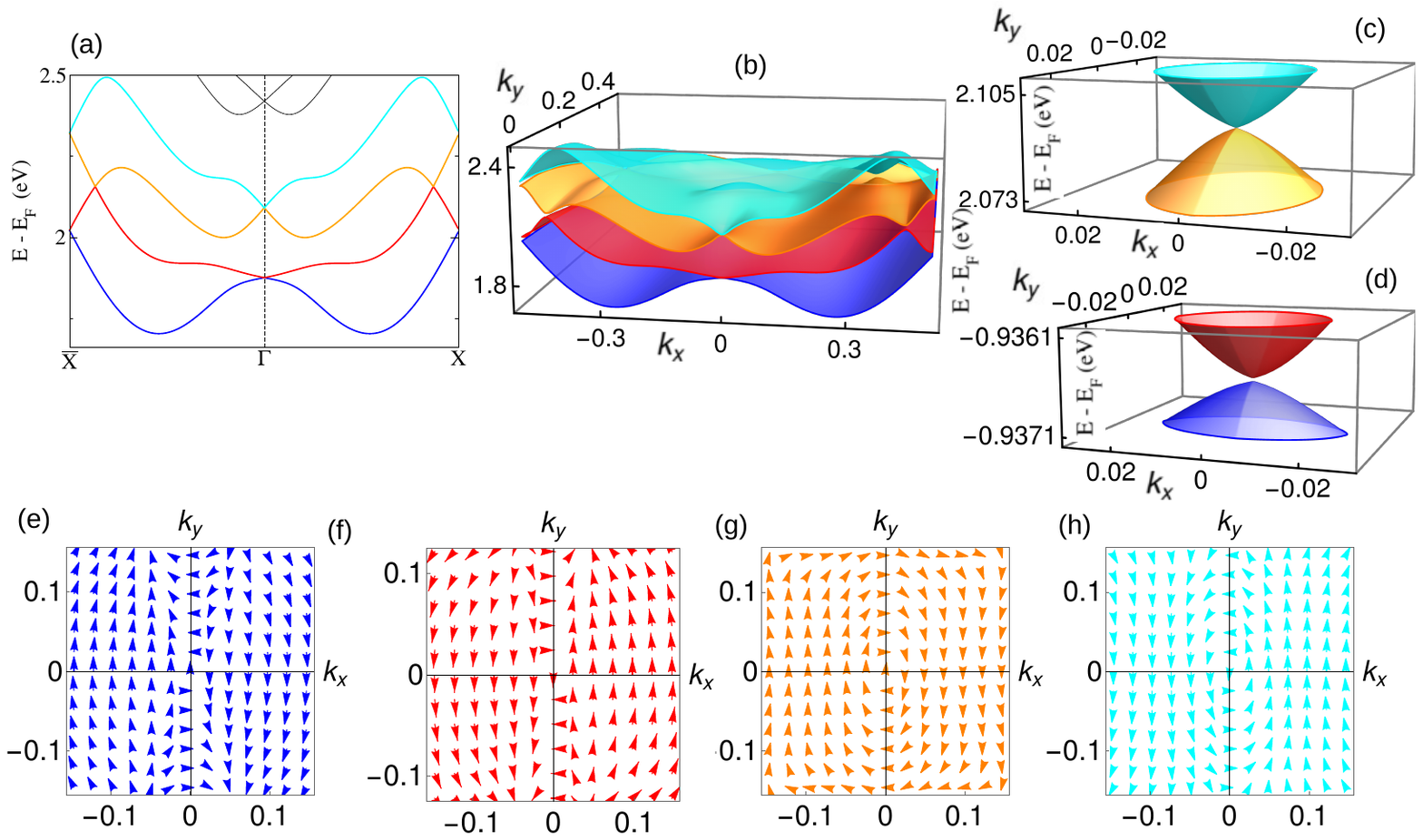}
    \caption{\label{fig:bulkwylebandspin}
(a) Hourglass band network formed by the lowest four conduction bands of bulk BiInO$_3$ along the high-symmetry path $\bar{X} \rightarrow \Gamma \rightarrow X$, where Kramers pairs exchange their partners as the path proceeds from $\Gamma$ to $X$.  
(b) Cross-sectional view of the corresponding 3D band structure.  
(c) and (d) Three-dimensional band structures focused on a narrow energy window around the $\Gamma$ point.  
(e)–(h) Persistent spin textures of the four conduction bands shown in (a), calculated using the first-principles DFT method. The colour of the spin corresponds to the colour of the band in (a).}

    \end{figure*}
    \begin{figure*}
        \centering
        \includegraphics[width=0.95\linewidth]{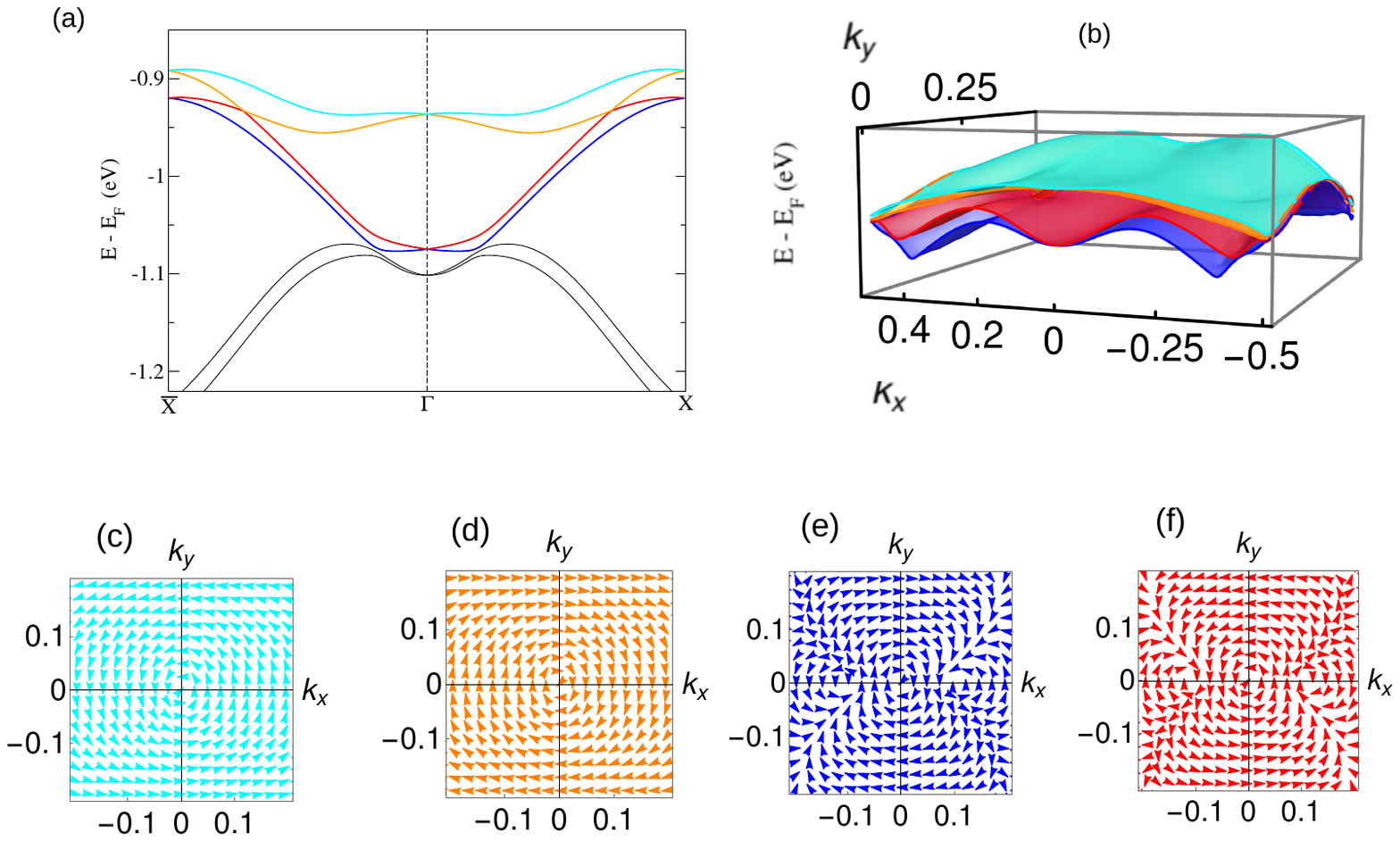}
        \caption{\label{fig:bulkwylebandspin}
(a) Hourglass band network formed by the highest four valence bands of bulk BiInO$_3$ along the high-symmetry path $\bar{X} \rightarrow \Gamma \rightarrow X$, where Kramers pairs exchange their partners as the path proceeds from $\Gamma$ to $X$.  
(b) Three-dimensional band structures focused on a narrow energy window around the $\Gamma$ point.  
(c)–(f) Persistent spin textures of the four valence bands shown in (a), calculated using the first-principles DFT method. The colour of the spin corresponds to the colour of the band in (a)}
        \label{fig:belowfermispin}
    \end{figure*}

\subsection{Electronic Structure of Bulk  BiInO$_3$}
A fascinating electronic structure has unfolded in our Density Functional Theory (DFT) calculations for both bulk and surface configurations of BiInO$_3$. Employing the Berry phase method, we computed the Electric polarisation, revealing a substantial value of approximately 12.4 $\mu$C/cm$^2$ along the [001] direction. The crystal structure manifests distinct polar planes, with the BiO and InO$_2$ layers exhibiting positive and negative charge accumulation, respectively, forming a serial combination of conventional capacitors along the z-direction. BiInO$_3$ emerges as a band insulator with an indirect band gap of 2.77 eV without SOI. Upon introducing SOI into our calculations, a notable reduction in the band gap to 2.55 eV occurs, accompanied by a profound band-splitting phenomenon. In specific directions, exhibit protected band-splitting due to mirror glide reflection symmetry, forming a Kramer nodal line \cite{Xie2021b}. In some other particular directions, this led to the emergence of Weyl points facilitated by the exchange of Kramer pairs. Our symmetry analysis is used to understand the topological band structure of BiInO$_3$, which will be explained in the coming sections.
  
\subsubsection{Symmetry analysis and Band structure}

(A) Consider the high-symmetry path $\bar{X}(-\pi, 0, 0) \rightarrow \Gamma(0, 0, 0) \rightarrow X(\pi, 0, 0)$, along which a general momentum point $(k_x, 0, 0)$ with $-\pi/a \leq k_x \leq \pi/a$ traces the trajectory. The little group along this path is composed of two key symmetries: the glide reflection $\left\{ m_{010} \,\middle|\, \tfrac{1}{2}~0~\tfrac{1}{2} \right\}$ (in the xz-plane) and the antiunitary symmetry $ \mathcal{T}\left\{ m_{100} \mid \frac{1}{2}~\frac{1}{2}~\frac{1}{2} \right\}$ (involving time-reversal and a glide along the yz-plane). These symmetries preserve $k_x$ along the path. As derived from \cref{eq:14}, the eigenvalues of the glide operator $\left\{ m_{010} \mid \frac{1}{2}~0~\frac{1}{2} \right\}$ vary along the path: at $\Gamma$ they are $(+i, -i)$, at $X$ they become $(+1, -1)$, and at $-\bar{X}$ they are $(-1, +1)$. Although $\zeta$ contains time-reversal symmetry, we find from the relation $ \mathcal{T}^2\{m_{100}^2|0~1~1\}$ that Kramers degeneracy is not enforced along this path. Consequently, the bands split away from the TRIM points, which are $\Gamma$ and $X$. However, at the TRIM points themselves, degeneracy is protected. This results in an hourglass-shaped band structure, where partner-switching within the symmetry-enforced eigenvalue pairs causes a band crossing at least once between $\Gamma$ and $X$\cite{PhysRevB.96.075110,PhysRevB.94.195109,liu2023dirac}. As illustrated in \cref{fig:comb}(c) and \cref{fig:comb}(d), such hourglass dispersion appears at 2.99 eV in the conduction band and at –0.094 eV in the valence band, respectively.

(B) Next, we consider the high-symmetry path $X(\pi, 0, 0) \rightarrow S(\pi, \pi, 0)$, where the momentum varies as $(\pi, k_y, 0)$ with $0 \leq k_y \leq \pi/b$. Along this path, the symmetries $\left\{ m_{100} \mid \frac{1}{2}~\frac{1}{2}~\frac{1}{2} \right\}$ and $ \mathcal{T}\left\{ m_{010} \,\middle|\, \tfrac{1}{2}~0~\tfrac{1}{2} \right\}$ form the little group, as $k_y$ remains invariant under both. From \cref{eq:12}, we find that $ \{m_{010}^2|1~0~0\} = -\identity$, indicating that time-reversal symmetry enforces Kramers degeneracy. Thus, the bands appear as pairs $(\psi, \mathcal{T}\left\{ m_{010} \mid \frac{1}{2}~0~\frac{1}{2} \right\} \psi)$, which are degenerate along the path. Since the crystal Hamiltonian commutes with these symmetries and is invariant under time reversal, the Bloch states are also eigenstates of $\left\{ m_{100} \mid \frac{1}{2}~\frac{1}{2}~\frac{1}{2} \right\}$. According to \cref{eq:13}, we can write the eigenvalue equations as $\left\{ m_{100} \mid \frac{1}{2}~\frac{1}{2}~\frac{1}{2} \right\} \psi^{\pm}(\vec{k}) = \pm \psi^{\pm}(\vec{k})$, and similarly, $\left\{ m_{100} \mid \frac{1}{2}~\frac{1}{2}~\frac{1}{2} \right\}(\mathcal{T}\left\{ m_{010} \mid \frac{1}{2}~0~\frac{1}{2} \right\} \psi^{\pm}(\vec{k})) = \pm (\mathcal{T}\left\{ m_{010} \mid \frac{1}{2}~0~\frac{1}{2} \right\} \psi^{\pm}(\vec{k}))$. As a result, the bands appear as two symmetry-protected Kramers doublets: $(\psi^{+}, \mathcal{T}\left\{ m_{010} \mid \frac{1}{2}~0~\frac{1}{2} \right\} \psi^{+})$ and $(\psi^{-}, \mathcal{T}\left\{ m_{010} \mid \frac{1}{2}~0~\frac{1}{2} \right\} \psi^{-})$.

(C) Consider now the high-symmetry path $S(\pi, \pi, 0) \rightarrow Y(0, \pi, 0)$, where the momentum vector is $(k_x, \pi, 0)$ with $0 \leq k_x \leq \pi/a$. Along this segment, the relevant symmetries are again $\left\{ m_{010} \,\middle|\, \tfrac{1}{2}~0~\tfrac{1}{2} \right\}$ and $\mathcal{T}\left\{ m_{100} \mid \frac{1}{2}~\frac{1}{2}~\frac{1}{2} \right\}$, which preserve $\vec{k}$ and hence form the little group. This is similar to the $\Gamma \rightarrow X$ segment discussed earlier. Here too, $\mathcal{T}^2\{m_{100}^2|0~1~1\}= -\identity$, enforcing Kramers degeneracy. The bands are therefore doubly degenerate along the path, forming Kramers pairs $(\psi(\vec{k}), \mathcal{T}\left\{ m_{010} \mid \frac{1}{2}~0~\frac{1}{2} \right\} \psi(\vec{k}))$ called Kramers nodel line\cite{xie2021kramers}.

(D) Lastly, we consider the path from $Y(0, \pi, 0) \rightarrow \Gamma(0, 0, 0)$, where the momentum varies as $(0, k_y, 0)$ with $0 \leq k_y \leq \pi/b$. The little group here includes $\left\{ m_{100} \mid \frac{1}{2}~\frac{1}{2}~\frac{1}{2} \right\}$ and $\mathcal{T}\left\{ m_{010} \mid \frac{1}{2}~0~\frac{1}{2} \right\}$, which again preserves $k_y$ along the path. This is similar in symmetry structure to the $X \rightarrow S$ path. However, in this case, \cref{eq:13} gives $ \mathcal{T}^2\{m_{010}^2|1~0~0\} = \identity$, which does not enforce Kramers degeneracy. The eigenvalues of $\left\{ m_{100} \mid \frac{1}{2}~\frac{1}{2}~\frac{1}{2} \right\}$ are $(+i, -i)$ at the $\Gamma$ point, and $(+1, -1)$ at $Y$. Because the eigenvalues change between the endpoints, the bands split along the path and must cross at least once due to partner switching. This again results in an hourglass-shaped dispersion along the $Y \rightarrow \Gamma$ direction.

\begin{figure*}[t!]
\label{fig:1}
    \centering
    \includegraphics[scale=0.63]{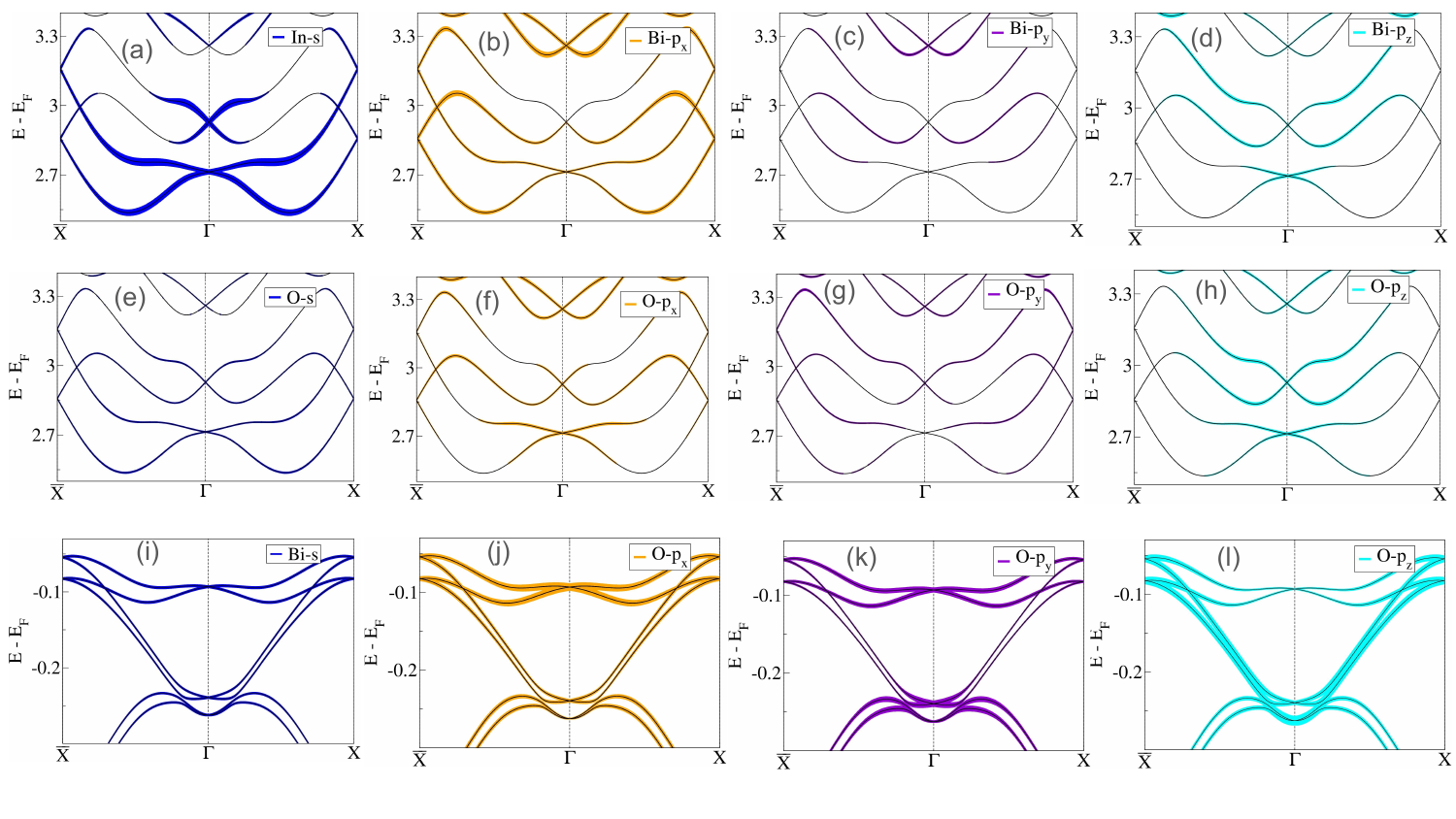}
    \caption{\label{fig:biinfat}
(a)–(h) Orbital-projected hourglass fermion band structures near the conduction band minimum. These bands exhibit a Persistent Spin Texture (PST) directed along the $k_y$-axis, primarily due to dominant contributions from In-$s$, Bi-$p_x$, and O-$p_y$ orbitals.  
(i)–(p) Orbital-projected hourglass fermion band structures near the Fermi energy around the $\Gamma$ point. These bands display Rashba-type spin texture, originating from the dominant contribution of O-$p_x$ and O-$p_y$ orbitals.
}

\end{figure*}

\begin{table*}[t!] 
\centering
\caption{
Transformation rules for the wave vector $\vec{k}$ and spin $\vec{\sigma}$ under the C$_{2v}$ point group symmetry operations at the $\Gamma(0, 0, 0)$ point in the Brillouin zone of BiInO$_3$. The wave vector $\vec{k}$ is defined relative to the high-symmetry $\Gamma$ point. Here, $\hat{\mathcal{K}}$ represents the complex conjugation operator.
\label{fig:c2v_transformations}}

\label{ch2tab1}
\resizebox{\linewidth}{!}{
\begin{tabular}{ccccc} 
\hline \hline
\multicolumn{3}{c}{} & \multicolumn{2}{c}{Invariant} \\ \hline
Symmetry &
Wave vector ($\vec{k}$) &
Spin ($\sigma$) &
Linear &
Cubic \\ 
\hline
$\mathcal{T}= i\sigma_y \hat{\mathcal{K}}$ & ($-k_x, -k_y, -k_z$) & ($-\sigma_x, -\sigma_y, -\sigma_z$) & $k_i \sigma_i, \, i=x,y,z$ & $k_i^3 \sigma_i, \, i=x,y,z$ \\
$\left\{ m_{100} \mid \frac{1}{2}~\frac{1}{2}~\frac{1}{2} \right\}=i\sigma_x$ & ($-k_x, k_y, k_z$) & ($\sigma_x, -\sigma_y, -\sigma_z$) & $k_x \sigma_y, k_x \sigma_z, k_y \sigma_x, k_z \sigma_x$ & $k_x^3 \sigma_y, k_x^3 \sigma_z, k_y^3 \sigma_x, k_z^3 \sigma_x$ \\ 
$\left\{ m_{010} \mid \frac{1}{2}~0~\frac{1}{2} \right\}=i\sigma_y$ & ($k_x, -k_y, k_z$) & ($-\sigma_x, \sigma_y, -\sigma_z$) & $k_x \sigma_y, k_y \sigma_x, k_y \sigma_z$ & $k_x^3 \sigma_y, k_y^3 \sigma_x, k_y^3 \sigma_z$ \\ 
$\left\{ 2_{001} \mid 0~0~\frac{1}{2} \right\} = i\sigma_z$ & ($-k_x, -k_y, k_z$) & ($-\sigma_x, -\sigma_y, \sigma_z$) & $k_z \sigma_z, k_i \sigma_i, \, i=x,y$ & $k_z^3 \sigma_z, k_i^3 \sigma_i, \, i=x,y$ \\ 
\hline \hline
\end{tabular}
}
\end{table*}

\subsubsection{$\vec{k} \cdot \vec{p}$ Hamiltonian}
The energy band splitting and their spin texture in BiInO$_3$ at $\Gamma$ point can be understood by writing the $\vec{k} \cdot \vec{p}$ Hamiltonian considering all the symmetries present. The point group symmetry at $\Gamma$ point is C$_{2v}$ characterized by two-fold rotation about z-axis ($\mathcal{C}_{2z}\left\{ 2_{001} \mid 0~0~0 \right\} $), reflection about xz-plane ($\mathcal{M}_y\left\{ m_{010} \mid 0~0~0 \right\}$) and reflection about yz-plane ($\mathcal{M}_x\left\{ m_{100} \mid 0~0~0 \right\}$). BiInO$_3$ is non-magnetic, thus preserving time-reversal symmetry ($\mathcal{T}$). The time-reversal symmetry reverses the direction of spinor $\vec{\sigma} \rightarrow -\vec{\sigma}$ and velocity $\vec{k}\rightarrow -\vec{k}$. Symmetries play a crucial role in understanding transformations of spin vectors on the Bloch sphere. The rotation of the Bloch sphere around an arbitrary axis, denoted by $\hat{n}$, is mathematically captured by the rotation matrix $R_{\hat{n}}(\theta)=e^{-i\frac{1}{2}\theta\hat{n} \cdot\vec{\sigma}}$ \cite{Bade1953,steane2013introduction}. This matrix transforms the Pauli spin matrices as $\sigma_i^{'}=R(\theta) \sigma_i R(\theta)^{-1}$, where $i$ can be x, y, or z. In the $\left\{ 2_{001} \mid 0~0~\frac{1}{2} \right\}$ symmetry, represented by $i\sigma_z$, the spinor on the XY-plane of the Bloch sphere undergoes a unique 
transformation: $(\sigma_x, \sigma_y, \sigma_z ) \rightarrow (-\sigma_x, -\sigma_y, \sigma_z )$. 
When a quantum spinor (such as a Pauli spin vector) is subjected to consecutive mirror reflections, the resulting operation is mathematically equivalent to a rotation. Specifically, two successive reflections about orthogonal mirror planes correspond to a $180^\circ$ rotation around the axis perpendicular to the plane defined by those mirrors. This concept, rooted in symmetry composition, plays a vital role in analysing spin textures. In spin systems, a mirror reflection about the XZ-plane ($\left\{ m_{010} \mid 0~0~0 \right\}$) corresponds to a $180^\circ$ rotation about the $y$-axis and is represented by the operator $i\sigma_y$. Under this transformation, the Pauli matrices transform as $(\sigma_x, \sigma_y, \sigma_z) \rightarrow (-\sigma_x, \sigma_y, -\sigma_z)$, preserving $\sigma_y$ and flipping $\sigma_x$ and $\sigma_z$. Similarly, a mirror reflection about the YZ-plane ($\left\{ m_{100} \mid 0~0~0 \right\}$)corresponds to a $180^\circ$ rotation about the $x$-axis, represented by $i\sigma_x$, which transforms the spin components as $(\sigma_x, \sigma_y, \sigma_z) \rightarrow (\sigma_x, -\sigma_y, -\sigma_z)$. These transformation rules are crucial for constructing effective $\vec{k} \cdot \vec{p}$ Hamiltonians in crystals with spin-orbit interaction, as they determine the allowed and forbidden spin terms under specific point-group symmetries. The complete transformation properties of wavevector $\vec{k}$ and spin $\vec{\sigma}$ under the relevant symmetry operations are summarised in Table~\ref{ch2tab1}.

Since BiInO$_3$ is a non-symorphic and non-magnetic material, spin splitting is observed due to spin-orbit interaction. The $\vec{k} \cdot \vec{p}$ Hamiltonian of the crystal is written using the theory of invariant \cite{book:924440,book:478467} as below,
\begin{equation}
    \mathcal{H}=\mathcal{H}_o + \mathcal{H}_{so}\label{eq:17}
\end{equation}
Where,
\begin{equation}
   \mathcal{H}_o = -\hbar \left( \frac{1}{2m_x} \frac{\partial^2}{\partial x^2} + \frac{1}{2m_y} \frac{\partial^2}{\partial y^2} \right)
      \label{eq:18}
\end{equation}
 The SOI term of the Hamiltonian H$_{so}$ is the sum of the product of Pauli matrices and the wavevector component transforming similarly under the symmetry present in the crystal. From the table, k$_x\rightarrow -k_x$ and $\sigma_y\rightarrow-\sigma_y$ under all the symmetries except M$_y$, where k$_x\rightarrow k_x$ and $\sigma_y\rightarrow\sigma_y$ on multiplying them, we get $k_x\sigma_y$, which is invariant under all the crystal symmetries and time-reversal symmetry. Thus, in a similar way of collecting the linear and cubic invariant terms under all the symmetries, we get $H_{so}$,
\begin{eqnarray}
    \mathcal{H}_{so}= &&\alpha_1k_x\sigma_y+\alpha_2k_y\sigma_x+\beta_1k_x^3\sigma_y+\beta_2k_y^3\sigma_x\nonumber\\
    &&+
    \beta_3k_x^2k_y\sigma_x+\beta_4k_y^2k_x\sigma_y 
   \label{eq:19}
\end{eqnarray}
The point group of BiInO$_3$ is C$_{2v}$, which belongs to the gyrotropic point group and allows only linear wavevector splitting\cite{Ganichev2014}. Considering only the linear terms, the Hamiltonian becomes.
\begin{equation}
    \mathcal{H}_{so}= \alpha_1k_x\sigma_y+\alpha_2k_y\sigma_x \label{eq:20}
\end{equation}
Rewriting the \cref{eq:20} as\cite{Tao2017},
\begin{equation}
    \mathcal{H}_{so}= \alpha_R (k_x\sigma_y+k_y\sigma_x)+\alpha_D(k_x\sigma_y-k_y\sigma_x)         \label{eq:21}
\end{equation}
where $\alpha_R=\dfrac{\alpha_1+\alpha_2}{2}$ and $\alpha_D =\dfrac{\alpha_1-\alpha_2}{2}$. 

The second part of the Hamiltonian is the Rashba Hamiltonian equal to $\alpha_R(\vec{k}\times\vec{\sigma})|_{z=0}$, and the first part of the Hamiltonian is the Dresselhaus Hamiltonian\cite{1001954}. $\alpha_R$ and $\alpha_D$ are the Rashba and Dresselhaus SOI constants, respectively.

The band-energy of \cref{eq:17} can be written as,
\begin{equation}
    \varepsilon^{\pm}(k)=\frac{k^2_x}{2m_x }+ \frac{k^2_y}{2m_y } \pm \sqrt{\alpha^2_1k^2_x+\alpha_2^2k_y^2}
    \label{eq:22}
\end{equation}
The eigenstate of the \cref{eq:17} can be written in spinor form as,
\begin{equation}
    \psi^{\pm}(k)=\frac{1}{\sqrt{2}}(\pm\xi \ket{\uparrow}+\ket{\downarrow})  \label{eq:23}
\end{equation}
Where,
\begin{equation}
    \xi =\frac{\sqrt{\alpha^2_1\cos^2\phi+\alpha^2_2\sin^2\phi}}{i\alpha_1\cos\phi+\alpha_2\sin\phi}    \label{24}
\end{equation}
$\ket{\uparrow}$ and $\ket{\downarrow}$ are the spinors for electrons spinning up and down, respectively, and $\phi$ is the angle between the spin and the k$_z$ axis.\\
The symmetry allowed expectation value of spin operator is $s^{\pm}=\frac{1}{2}\bra{\psi^{\pm}(k)}\sigma\ket{\psi^{\pm}(k)}$ results,
\begin{equation}
   \langle s_x\rangle^{\pm}=\frac{1}{2}\left(\frac{\pm\alpha_2\sin\phi}{\sqrt{\alpha^2_1\cos^2\phi+\alpha^2_2\sin^2\phi}} \right)   \label{eq:25}
\end{equation}
\begin{equation}
  \langle s_y\rangle^{\pm}=\frac{1}{2}\left(\frac{\pm\alpha_1\cos\phi}{\sqrt{\alpha^2_1\cos^2\phi+\alpha^2_2\sin^2\phi}} \right)     \label{eq:26}
\end{equation}
\begin{equation}
  \langle s_z\rangle^{\pm}= 0  \label{eq:27}
\end{equation}

\begin{table}[t]
\caption{\label{tab:Table2}Character table for the $C_{2v}$ point group.}
\centering
\begin{ruledtabular}
\begin{tabular}{cccccccc}
IR & \identity & $\mathcal{C}_2$ & $\mathcal{M}_x$ & $\mathcal{M}_y$ & Linear & Quadratic & Rotation \\
\hline
$A_1$ & 1 & 1 & 1 & 1 & $z$ & $x^2$, $y^2$, $z^2$ & \\
$A_2$ & 1 & 1 & $-1$ & $-1$ &  & $xy$ & $\sigma_z$ \\
$B_1$ & 1 & $-1$ & 1 & $-1$ & $x$ & $xz$ & $\sigma_y$ \\
$B_2$ & 1 & $-1$ & $-1$ & 1 & $y$ & $yz$ & $\sigma_x$ \\
\end{tabular}
\end{ruledtabular}
\end{table}

\subsection{Spin Textue}
 In our study, we utilised the DFT, $\vec{k} \cdot \vec{p}$ model, and group theory to analyse the spin texture of the energy band near the $\Gamma$ point in momentum space. We observed that two pairs of bands between $\Gamma$ and X point shared Kramer partners, influenced by the symmetries along the high symmetry path, which glue up the band between the high symmetry points. The band's double degeneracy at high symmetry points $\Gamma$ and X is due to time reversal symmetry. From \cref{fig:bulkwylebandspin}, we can see the up-down-down-up spin arrangement of the hourglass bands. The direction of the spin arrangement of the hourglass is governed by the symmetry at the high symmetry path. Specifically, at $\Gamma$ point both mirror plane $\left\{ m_{010} \mid \frac{1}{2}~0~\frac{1}{2} \right\}$ and $\left\{ m_{100} \mid \frac{1}{2}~\frac{1}{2}~\frac{1}{2} \right\}$ intersect to each-other. Both $\left\{ m_{010} \mid \frac{1}{2}~0~\frac{1}{2} \right\}$ and $\left\{ m_{100} \mid \frac{1}{2}~\frac{1}{2}~\frac{1}{2} \right\}$ mirror symmetries keep the $\Gamma$ point invariant, and the crystal Hamiltonian commutes with both these symmetries. As a result, the direction of the spin at the $\Gamma$ point is determined by this symmetry.  $\left\{ m_{010} \mid \frac{1}{2}~0~\frac{1}{2} \right\}$ leaves $\sigma_y$ invariant while flipping the direction of $\sigma_x$ and $\sigma_z$. Consequently, $\langle \sigma_x\rangle$ and $\langle \sigma_z\rangle$ are zero, while $\langle \sigma_y\rangle$ non-zero. Similarly, for $\left\{ m_{010} \mid \frac{1}{2}~0~\frac{1}{2} \right\}$ $\langle \sigma_y\rangle$ and $\langle \sigma_z\rangle$ are zero. In contrast, $\langle 
 \sigma_x\rangle$ is non-zero. $\langle \sigma_y\rangle$ and $\langle \sigma_y\rangle$ are calculated in \cref{eq:25,eq:26,eq:27} using $\vec{k} \cdot \vec{p}$ hamiltonian.
 
 \subsubsection{\textbf{Spin Texture of Valance band hourglass fermion network at $\Gamma$ point}} 
 
After fitting the $\vec{k} \cdot \vec{p}$ model to the DFT results around point C of \ref{fig:comb}(d), we determined that $\alpha_1=0.06$ and $\alpha_2=-0.06$, yielding Dresselhaus SOI constant $\alpha_D =0$ and Rashba SOI constant $\alpha_R=0.12$. Consequently, the spin of the hourglass fermion exhibits a Rashba spin texture. Similarly, at point D of \cref{fig:comb}(d), we obtained $\alpha_1=0.01$ and $\alpha_2=-0.01$, resulting in Dresselhaus SOI constant $\alpha_D=0$ and Rashba SOI constant $\alpha_R=0.02$. This illustrates that Rashba SOI effectively suppresses Dresselhaus SOI,
leading to a Rashba spin texture. The corresponding spin texture is shown in \cref{fig:belowfermispin}
To understand the spin texture, we examine the invariant of the basis function of the orbitals contributing to the band \cite{Bandyopadhyay2021}. The point group at the $\Gamma$ point is $\mathcal{C}_{2v}$, with its character table displayed in \cref{tab:Table2}. In our analysis of the orbital contribution to the energy band, we plotted the orbital projected band for the hourglass fermions in \cref{fig:biinfat}. We observe that O-p$_x$ and O-p$_y$ orbitals make equal contributions, surpassing those from O-p$_z$.
Referring to \cref{tab:Table2}, we note that O-p$_x$ and $\sigma_y$ follow the B$_1$ representation, which is not invariant under $\left\{ 2_{001} \mid 0~0~0\right\}$ and $\left\{ m_{010} \mid 0~0~0 \right\}$, but is invariant under $\left\{ m_{100} \mid 0~0~0 \right\}$. Conversely, O-p$_y$ and $\sigma_x$ follow the B$_2$ representation, which is not invariant under $\left\{ 2_{001} \mid 0~0~0 \right\}$ and $\left\{ m_{100} \mid 0~0~0 \right\}$, but invariant under $\left\{ m_{010} \mid 0~0~0 \right\}$. Given the nearly equal contributions of O-p$_x$ and O-p$_y$, the spin direction is governed by the expectation of both $\sigma_x$ and $\sigma_y$. This results in a Rashba spin texture, consistent with the DFT-calculated spin texture and the $\vec{k} \cdot \vec{p}$ model.

\subsubsection{\textbf{Spin texture of conduction band hourglass fermion network at $\Gamma$ point}}
The conduction band minimum lies between the $\Gamma$ and X points at 2.53eV. This minimum is one of the bands among a set of four bands contributing to the formation of the Hourglass band dispersion. We calculated the spin texture for the four bands, as shown in \cref{fig:bulkwylebandspin}.
Upon fitting the $\vec{k} \cdot \vec{p}$ model with DFT results at point A of \cref{fig:comb}(c), we determined $\alpha_1 = 0.195$ and $\alpha_2 = -0.01$. We calculated the Rashba and Dresselhaus spin-orbit interaction (SOI) constants from these values, yielding $\alpha_R = 0.196$ for Rashba SOI and $\alpha_D = 0.194$ for Dresselhaus SOI. Remarkably, these constants are nearly equal, resulting in a momentum-independent spin texture point along the k$_y$ direction, referred to as Persistent Spin Texture (PST). Similarly, at point B of \cref{fig:comb}(c), the equality of $\alpha_1$ and $\alpha_2$ values also gives rise to PST. Consequently, the spin polarisation of the Hourglass Fermion along the $\Gamma$ to X direction is arranged as up-down-down-up along the k$_y$ direction.
On the other hand, the spin texture of the valence band around the $\Gamma$ point at points C and D exhibits a Rashba spin texture, with equal contributions from $\sigma_x$ and $\sigma_y$, consistent with symmetry analysis, DFT, and $\vec{k} \cdot \vec{p}$ model predictions. However, the spin texture of the conduction band differs significantly from that of the valence band, despite both bands sharing the same C$_{2v}$ point group present at the $\Gamma$ point. Given that the spin texture points along the k$_y$ direction, the expectation value of $\sigma_x$ is nearly zero. Understanding a clear picture of the spin texture necessitates examining the competition of orbital characters in the bands. \Cref{fig:biinfat} illustrates the orbital-projected bands, revealing that in hourglass fermions, the dominant orbital characters originate from In-s, Bi-p$_x$, and O-p$_x$ orbitals, compared to the contributions from Bi-p$_y$ and O-p$_y$. The p$_x$ orbital of Bi and O follows the B$_1$ representation, while the p$_y$ orbital of Bi and O follows the B$_2$ representation. Notably, the rotational part $\sigma_y$ follows the B$_1$ representation, whereas $\sigma_x$ follows the B$_2$ representation. As the linear basis function following the B$_1$ representation dominates over that following the B$_2$ representation, $\sigma_y$ surpasses $\sigma_x$, resulting in the spins of hourglass fermions aligning along the k$_y$ direction.

\section{\label{sec:conc}Conclusion} In this research, we investigated the electronic structure of semiconducting bulk BIO within the frame-
work of density-functional theory, including spin-orbit interaction. The study provides insights into how the non-symmorphic crystal symmetries of BIO, such as glide reflection symmetry, impact its energy band structure. BIO exhibits an hourglass fermion attributed to its non-symmorphic crystal symmetries. The interplay between the SOI and point group symmetry at the $\Gamma$ point results in unique spin-momentum locking in the band near the Fermi energy, and results in Rashba spin texture at $\Gamma$ point below the Fermi energy and PSH above the Fermi energy at $\Gamma$ high symmetry point. Though both bands share the same point group symmetry, they feature two different types of spin-texture due to the competition of p-orbital character of constituent atoms in the bands. Below the $\Gamma$-point we find Rashba spin-texture due to equal contribution coming from O-p$_x$ and O-p$_y$, and above $\Gamma$-point there is PSH along the y-direction as Bi-p$_x$ and O-p$_x$ is dominating. The direction of spin-texture is also calculated using $\vec{k} \cdot \vec{p}$ method, which matches with the DFT result. The unique spin textures in BIO, driven by symmetry-enforced mechanisms, hold significant potential for spintronics applications. The PST in the conduction band is particularly advantageous for developing reliable and efficient spin field-effect transistors (spin-FETs) by minimising spin relaxation challenges\cite{Datta1990,Dyakonov2017}.

\begin{acknowledgments}
RK acknowledges UGC, India, for a research fellowship through grant number 1500/(CSIR-UGC NET JUNE 2017). NG acknowledges financial support from SERB, India, through grant number CRG/2021/005320. The use of high-performance computing facilities at IISER Bhopal is gratefully acknowledged.
\end{acknowledgments}

\bibliography{library1}
\end{document}